\def\year{2018}\relax
\documentclass[letterpaper]{article} 
\usepackage{aaai18}  
\usepackage{times}  
\usepackage{helvet}  
\usepackage{courier}  
\usepackage{url}  
\usepackage{graphicx}  

\usepackage{multirow}
\usepackage{subfigure}

\usepackage{color}
\usepackage{booktabs}

\usepackage[colorinlistoftodos]{todonotes}

\begin{document}
%
\title{Learning Graph Topological Features via GAN}

\author{
Weiyi Liu\textsuperscript{1,2},
Hal Cooper\textsuperscript{3},
Min Hwan Oh\textsuperscript{3},
Sailung Yeung\textsuperscript{4},
Pin-Yu Chen\textsuperscript{2}
Toyotaro Suzumura\textsuperscript{2}
Lingli Chen\textsuperscript{1}\\
\textsuperscript{1}{University of Electronic Science and Technology of China},
\textsuperscript{2}{IBM Watson Research Center},
\textsuperscript{3}{Columbia University},
\textsuperscript{4}{Boston University}\\
weiyiliu@us.ibm.com,
hal.cooper@columbia.edu, 
m.oh@columbia.edu\\
yeungsl@bu.edu,
pin-yu.chen@ibm.com,
suzumura@acm.org,
lingli324@std.uestc.edu.cn
}


\maketitle
\begin{abstract}
Inspired by the generation power of generative adversarial networks (GANs) in image domains, we introduce a novel hierarchical architecture for learning characteristic topological features from a single arbitrary input graph via GANs. The hierarchical architecture consisting of  multiple GANs preserves both local and global topological features and automatically partitions the input graph into representative “stages” for feature learning. The stages facilitate reconstruction and can be used as indicators of the importance of the associated topological structures. Experiments show that our method produces subgraphs retaining a wide range of topological features, even in early reconstruction stages (unlike a single GAN, which cannot easily identify such features, let alone reconstruct the original graph). This paper is firstline research on combining the use of GANs and graph topological analysis. 
\end{abstract}

\section{Introduction}\label{sec:introduction}
Graphs have great versatility, able to represent complex systems with diverse relationships between objects and data. With the rise of social networking, and the importance of relational properties to the ``big data'' phenomenon, it has become increasingly important to develop ways to automatically identify key structures present in graph data. Identification of such structures is crucial in understanding how a social network forms, or in making predictions about future network behavior. 
To this end, a large number of graph analysis methods have been proposed to analyze the topology of the target network at the node \cite{muppidi2016survey}, community \cite{fortunato2010community,martinez2016survey}, and global levels \cite{wu2017evaluation}, and perform certain tasks. 

Unfortunately, each level of analysis is greatly influenced by network topology, and so far no algorithm can be adapted to work effectively for arbitrary network structures. Modularity-based community detection \cite{xiang2016local} works well for networks with separate clusters, whereas edge-based methods \cite{delis2016scalable} are suited to dense networks. Similarly, when performing graph sampling, Random Walk (RW) is suitable for sampling paths \cite{leskovec2006sampling}, whereas Forrest Fire (FF) is useful for sampling clusters \cite{leskovec2005graphs}. When it comes to graph generation, Watts-Strogatz (WS) graph models \cite{watts1998collective} can gererate graphs with small world features, whereas Barabási-Albert (BA) graph models \cite{barabasi1999emergence} simulate super hubs and regular nodes according to the scale-free features of the network.

However, real-world networks typically have multiple topological features. Considering real-world networks also introduces another issue that traditional graph analysis methods struggle with; we may only have a single instance of a graph (e.g. the transaction graph for a particular bank), making it difficult to identify the key topological properties in the first place. In particular, we are interested in both ``local topological features’’ (such as the presence of subgraph structures like triangles) and ``global topological features'' such as degree distribution.

\begin{figure}
\includegraphics[width=\linewidth]{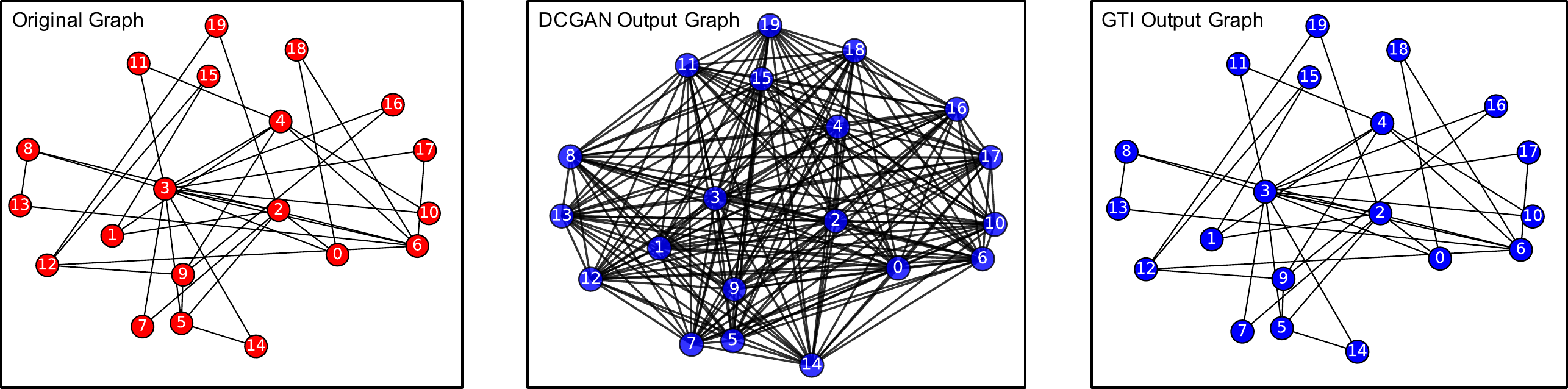}
  \caption{How GTI recovers the original graph while naive GAN methods do not}
\label{fig-compare}
\end{figure}

In this paper we propose an unsupervised method, the Graph Topology Interpolater (GTI), to facilitate graph analysis without regard for these issues. GTI is a novel approach combining techniques from graph analysis and GAN based image processing techniques. In Figure \ref{fig-compare}, we demonstrate that naively inputting the full graph (here, a 20 node BA network \cite{barabasi1999emergence}) into a standard GAN representation (the DCGAN \cite{radford2015unsupervised}) is unsuccessful; such a GAN structure is unable to learn to reproduce the original graph, instead getting stuck in undesirable local minima. 

Therefore, instead of directly analyzing the entire topology of a graph, GTI first divides the graph into several hierarchical layers. A hierarchical view of a graph can split the graph by local and global topological features, giving a better understanding of the graph \cite{watts2002identity}. As different layers have different topological features, GTI uses separate GANs to learn each layer and the associated features. By leveraging GANs renowned feature identification \cite{goodfellow2014generative,goodfellow2016nips} on each layer, GTI has the ability to automatically capture arbitrary topological features from a single input graph. 



In addition to learning topological features from the input graph, the GTI method defines a reconstruction process for reproducing the original graph via a series of reconstruction stages (the number of which is automatically learned during training). As stages are ranked in order of their contribution to the full topology of the original graph, early stages can be used as an indicator of the most important topological features. Our focus in this initial work is on the method itself and demonstrating our ability to learn these important features quickly (via demonstrating the retention of identifiable structures and comparisons to graph sampling methods).

\section{Method}\label{sec:method}
\begin{figure*}[h]
\centering
\scalebox{0.8}{
	\includegraphics[width=1\textwidth]{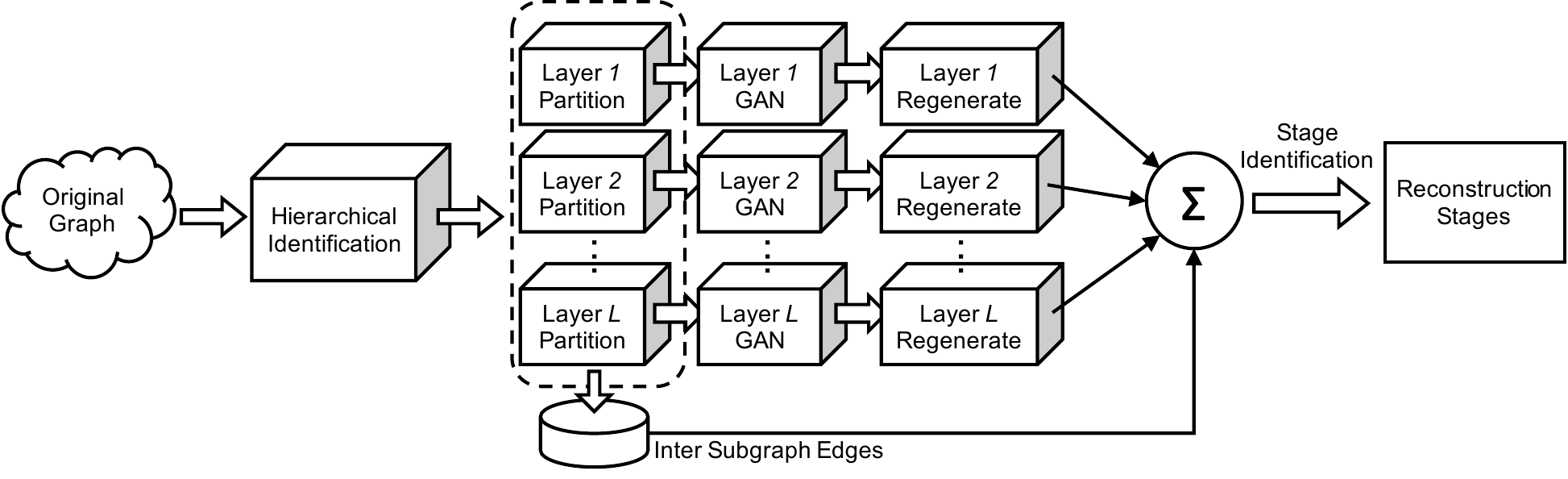}
}
	\caption{Work flow for graph topology interpolater (GTI)}
	
	\label{fig-HGG-flow}

\end{figure*}

In this section, we demonstrate the work flow of our Graph Topology Interpolater (GTI) (Figure \ref{fig-HGG-flow}), with a particular focus on the GAN, Sum-up and Stage Identification modules. At a high level, the GTI method takes an input graph, learns it's hierarchical layers, trains a separate GAN on each layer, and autonomously combines their output to reconstruct stages of the graph. Here we give a brief overview of each module.

\textbf{Hierarchical Identification Module:} This module detects the hierarchical structure of the original graph using the Louvain hierarchical community detection method \cite{blondel2008fast}, denoting the number of layers as $L$. The number of communities in each layer is used as a criterion for how many subgraphs a layer should pass to the next module.

\textbf{Layer Partition Module:} The main purpose of this module is to partition a given layer into $M$ non-overlapping subgraphs, where $M$ is the number of communities. The reason why we do not use the learned communities from the Louvain method is that we cannot constrain the size of any community. We instead balance the communities into fixed size subgraphs using the METIS approach \cite{karypis1995metis}.


\textbf{Layer GAN Module:} Here we apply the GAN methodology to graph analysis. Rather than directly using one GAN to learn the whole graph, we use different GANs to learn features for each layer separately. If we use a single GAN to learn features for the whole graph, some topological features may be diluted or even ignored. For more detail see Section GAN\ref{subsec:GAN}.

\textbf{Layer Regenerate Module:} Here, for a given layer, the corresponding GAN has learned all the properties of each subgraph, meaning we can use the generator in this GAN to regenerate the topology of the layer by generating $M$ subgraphs of $k$ nodes.
Note that this reconstruction only restores edges within each non-overlapping subgraph, and does not include edges between subgraphs.

\textbf{All Layer Sum-up Module:} This module outputs a weighted reconstructed graph by summing up all reconstructed layers along with the edges between subgraphs that were not considered in the Regenerate Module. The ``weight'' of each edge in this module represents its importance to the reconstruction. Indeed, we rely upon these weights to identify the reconstruction stages for the original graph. For more details, see Section Sum \ref{subsec:sum}.
%
%

\textbf{Stage Identification Module:} By analyzing the weighted adjacency matrix of the Sum-up Module, we extract stages for the graph. These stages can be interpreted as steps for graph reconstruction process. See Section Stage\ref{subsec:stage} for details.

\subsection{Layer GAN Module}\label{subsec:GAN}

\begin{figure}[h]
	\centering
    \subfigure[Generator Architecture]
    {
    	\label{fig-gen}
    	\includegraphics[width=0.45\textwidth]{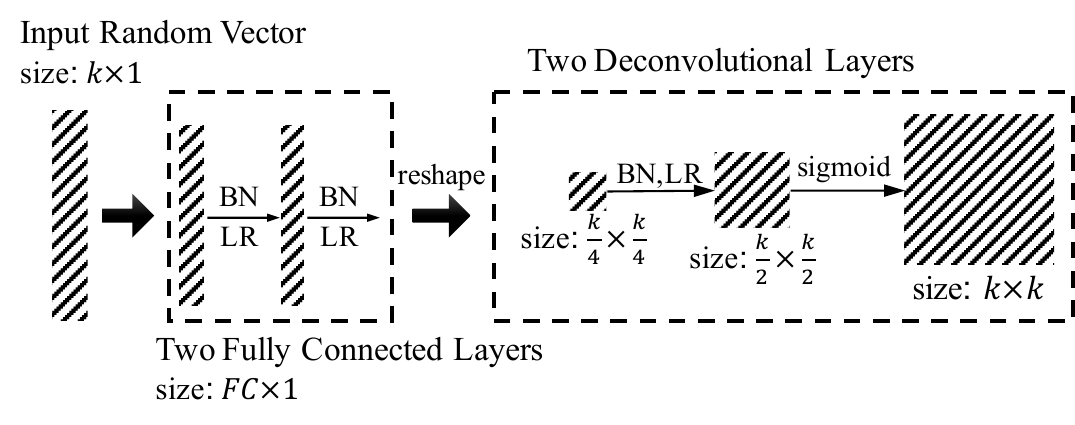}
    }
    \subfigure[Discriminator Architecture]
    {
        \label{fig-disc}
    	\includegraphics[width=0.45\textwidth]{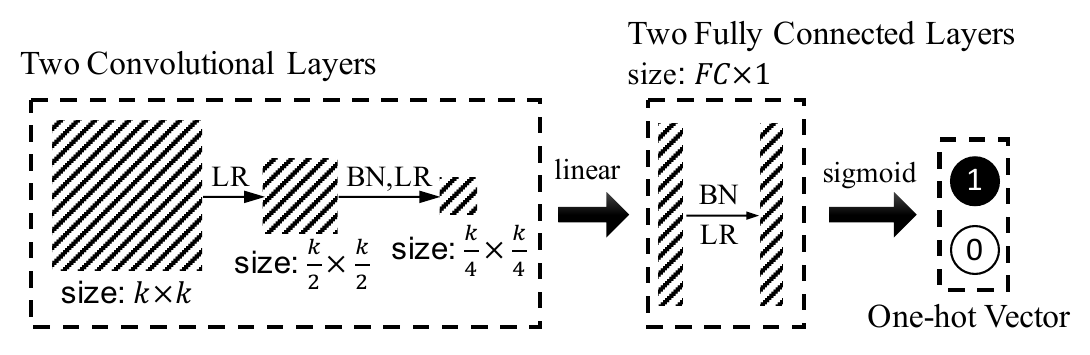}
    }
    \caption{Generator and discriminator architecture}
    \label{fig-gen-disc}
\end{figure}

Figure \ref{fig-gen} and Figure \ref{fig-disc} show the architectures for the generator and discriminator of the GAN. Where the generator is a deconvolutional neural network with the purpose of restoring a $k\times k$ adjacency matrix from the standard uniform distribution, the discriminator is instead a CNN whose purpose is to estimate if the input adjacency matrix is from a real dataset or from a generator. Here, $BN$ is short for batch normalization which is used instead of max pooling because max pooling selects the maximum value in the feature map and ignores other values, whereas $BN$ will synthesize all available information. $LR$ stands for the leaky ReLU active function ($LR=max(x, 0.2\times x)$) which we use as the value $0$ has a specific meaning for adjacency matrices. In addition, $k$ represents the size of a subgraph, and $FC$ the length of a fully connected layer. We set the stride for the convolutional/deconvolutional layers to be $2$. We adopt the same loss function and optimization strategy (1000 iterations of ADAM \cite{kingma2014adam} with a learning rate of 0.0002) used in DCGAN \cite{radford2015unsupervised}.



\subsection{Sum-up Module} \label{subsec:sum}
In this module, we use a linear function (see Equation \ref{eq-sum}) to add the graphs from all layers together. $re_G$ stands for the reconstructed adjacency matrix (with input from all layers), $G'_i, i \in L$ represents the reconstructed adjacency matrix for each layer (with $G$ representing the full original graph with $N$ nodes), $E$ refers to all the inter-subgraph (community) edges identified by the Louvain method from each hierarchy, and $b$ represents a bias. Note that while each layer of the reconstruction may lose certain edge information, summing up the  hierarchical layers along with $E$ will have the ability to reconstruct the entire graph.
\begin{equation}
re_G = \sum_{i=1}^L w_i G'_i + wE + b \label{eq-sum}
\end{equation}
To obtain the weight $w$ for each layer and the bias $b$, we use Equation \ref{eq-sum-loss} as the loss function (where we add $\epsilon=10^{-6}$ to avoid taking $log(0)$ or division by 0), using 500 iterations of SGD with learning rate 0.1 to minimize this loss function, and find suitable parameters. We note that Equation \ref{eq-sum-loss} is similar to  a $KL$ divergence, though of course $re_G$ and $G$ are not probability distributions.
\begin{equation}
\textrm{ Loss }(re_{ G },G)=\sum _{ i\in 1\cdots N^{ 2 } }^{  }{ vec(G+\epsilon )_i\cdot log\frac { vec(G+\epsilon )_i }{ vec(re_{ G }+\epsilon )_i }  }  
\label{eq-sum-loss}
\end{equation}

\subsection{Stage Identification} \label{subsec:stage}
After the above calculations, we obtain the weighted adjacency matrix $re_G$, where weights on edges represent how each edge contributes to the entire topology. Clearly, different weights represent different degrees of contribution to the topology. Therefore, according to these weights, we can divide the network into several stages, with each stage representing a collection of edges greater than a certain weight. We introduce the concept of a ``cut-value'' to turn $re_G$ into a binary adjacency matrix.
We observe that many edges in $re_G$ share the same weight, which implies these edges share the same importance. Furthermore, the number of unique weights can define different reconstruction stages, with the most important set of edges sharing the highest weight. 
Each stage will include edges with weights greater than or equal to the corresponding weight of that stage. Hence, we define an ordering of stages by decreasing weight, giving insight on how to reconstruct the original graph in terms of edge importance. We denote the $i$th largest unique weight-value as $CV_i$ (for ``cut value'') and thereby denote the stages as in Equation \ref{eq:stage_def} (an element-wise product), where $I[w\geq CV_i]$ is an indicator function for each weight being equal or larger than the $CV_i$.
\begin{equation}
re_G^i = re_G I[w\geq CV_i]
\label{eq:stage_def}
\end{equation}


In Section Evaluation\ref{sec:evaluation}, we use synthetic and real networks to show that each stage preserves identifiable topological features of the original graph during the graph reconstruction process. As each stage contains a subset of the original graphs edges, we can interpret each stage as a sub-sampling of the original graph. This allows us to compare with prominent graph sampling methodologies to emphasize our ability to retain important topological features.
\section{Related Work} \label{sec:works}
The development of deep learning and the growing maturity of graph topology analysis has led to more attention on the ability to use the former for the latter \cite{li2015gated}. A number of supervised and semi-supervised learning method have been developed for graph analysis. A particular focus is on the use of CNNs \cite{bruna2014spectral,henaff2015deep,duvenaud2015convolutional,defferrard2016convolutional}.  These new methods have shown promising results on their respective tasks in comparison to traditional graph analysis methods (such as kernel-based methods, graph-based regularization techniques, etc). Kipf et al. has discussed the above methods in detail \cite{defferrard2016convolutional,kipf2016semi}, pointing out the strengths and drawbacks of various approaches. 
Recently, Sahar et al. \cite{tavakolilearning} have proposed a naive method to generate the graph topology using GAN, by randomly perturbing the input graph 10,000 times, and feed these graphs into a DCGAN. It simply treats the adjacency matrix of a graph as an image, and uses random permutation to generate enough samples to train a DCGAN, without taking any topological information of the graph into consideration.

A key difference between GTI and other methods is that GTI is an unsupervised learning tool (facilitated by the use of GANs), that leverages the hierarchical structure of a graph. GTI can automatically capture both local and global topological features of a network. To the best of the authors' knowledge, this is the first unsupervised method in such manner.

Since GANs were first introduced \cite{goodfellow2014generative} in 2014, its theory and application has expanded greatly. Many advances in training methods \cite{denton2015deep,chen2016infogan,zhao2016energy,nowozin2016f} have been proposed in recent years, and this has facilitated their use in a number of applicaitons.
For example, GAN has been used for artwork synthesis \cite{tan2017artgan}, text classification \cite{miyato2016adversarial}, image-to-image translation \cite{yi2017dualgan}, imitation of driver behavior \cite{kuefler2017imitating}, identification of cancers \cite{kohl2017adversarial}, e.t.c. The GTI method expands the use of GANs into the graph topology analysis area. 

\section{Evaluation}\label{sec:evaluation}
All experiments in this paper were conducted locally on CPU using a Mac Book Pro with an Intel Core i7 2.5GHz processor and 16GB of 1600MHz RAM.
Though this limits the size of our experiments in this preliminary work, the extensive GAN literature (see Section Works\ref{sec:works}) and the ability to parallelize GAN training based on hierarchical layers suggests that our method can be efficiently scaled to much larger systems.

\subsection{Datasets}
We use a combination of synthetic and real datasets. Through the use of synthetic datasets with particular topological properties, we are able to demonstrate the retention of these easily identifiable properties across the reconstruction stages. Of course, in real-world applications we do not know the important topological structures a priori, and so also demonstrate our method on a number of real-world datasets of varying sizes.

We use the ER graph model \cite{ER}, the BA graph model \cite{barabasi1999emergence}, the WS graph model \cite{watts2002identity} and the Kronecker graph model \cite{leskovec2010kronecker} to generate our synthetic graphs. The varying sizes of our synthetic graphs (as well as our real-world datasets) are outlined in Table \ref{tab:Table-datasets}. 

For real datasets, we use data available from the Stanford Network Analysis Project (SNAP) \cite{SNAP}. In particular, we use the Facebook network, the wiki-Vote network, and the P2P-Gnutella network. The Facebook \cite{FB} dataset consists of ``friends lists'', collected from survey participants according to the connections between user-accounts on the online social network. It includes node features, circles, and ego networks; all of which has been anonymized by replacing the Facebook-internal ids. Wiki-vote \cite{wiki} is a voting network (who votes for whom etc) that is used by Wikipedia to elect page administrators; P2P-Gnutella \cite{P2P} is a peer-to-peer file-sharing network: Nodes represent hosts in the Gnutella network topology, with edges representing connections between the hosts. RoadNet \cite{roadNet} is the road network of Pennsylvania. Intersections and endpoints are represented by nodes, and the roads connecting them are edges. As this graph is of a size prohibitive to the compute resources used in this preliminary work, we choose a connected component of appropriate size (note that the full network is not connected because nodes may be connected in the real-world by roads outside of Pennsylvania).

\begin{table*}[t]
  \caption{Size of original datasets, and corresponding reconstruction stages}
  \label{tab:Table-datasets}
  \centering
  \scalebox{1}{
  \begin{tabular}{lllll}
    \toprule
    Graph 	& 	\# Nodes 	&	\# Edges & \# Stages & Retained edge percentage for ordered stages (\%)\\
    \toprule
    BA		&	500		&	996	  & 7	&19.48, 26.31, 36.04, 39.36, 41.57, 57.43, 100\\
    ER		&	500		&	25103 & 4	&4.32, 21.73, 94.91, 100\\
    Kronecker&	2178	&	25103 & 10	&87.77, 88.65, 91.76, 91.89, 92.47, 93.32, 96.06, 97.05, 98.57, 100\\
    WS		&	500		&	500	  & 7	&11.20, 11.40, 16.00, 18.00, 54.60, 97.80, 100\\
    \midrule
    Facebook&	4039	&	88234 & 7	&52.28, 83.33, 87.49, 91.41, 90.31, 91.95, 100\\
    Wiki-Vote&	7115	&	103689& 4	&58.31, 73.79, 85.60, 100\\
    \centering RoadNet	&	5371	&	7590  & 12	&0.62, 3.87, 26.64, 27.98, 31.79, 32.42, 34.22, 34.65, 34.80, 64.06, 76.81, 100\\
    P2P		&	3334	&	6627  & 7	&49.04, 53.90, 70.32, 87.54, 88.40, 89.65, 100\\
    \bottomrule
  \end{tabular}}
\end{table*}

\subsection{Local Topological Features}
\begin{figure}
\includegraphics[width=\linewidth]{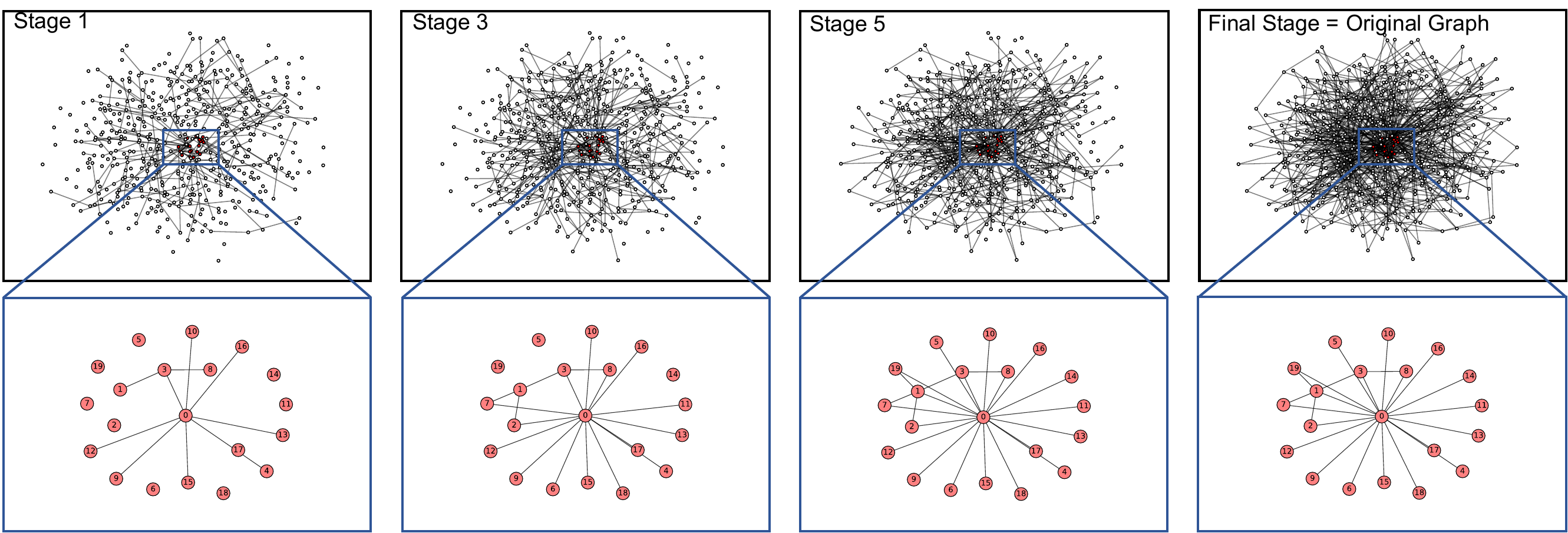}
  \caption{The topology of original graph and corresponding stages of BA network.}
\label{evolve-synthetic}
\end{figure}

\begin{figure}
\includegraphics[width=\linewidth]{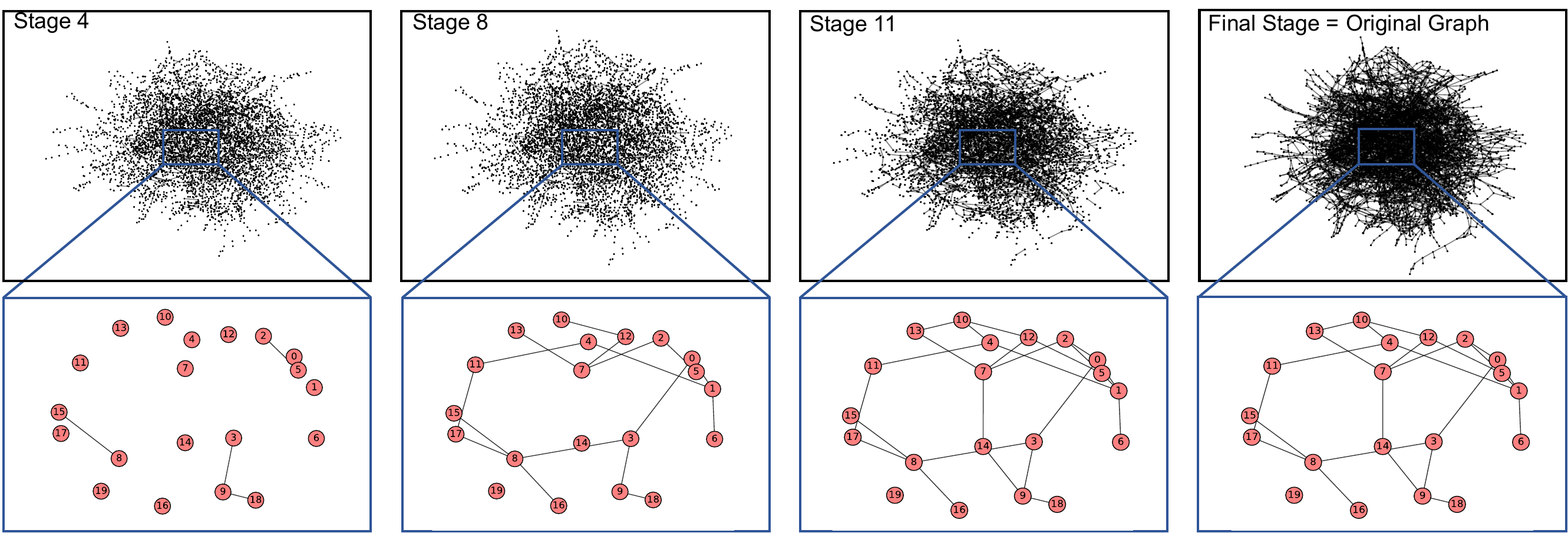}
  \caption{The topology of original graph and corresponding stages of road network.}
\label{evolve-real}
\end{figure}

As discussed in Section Method\ref{sec:method}, GTI automatically ranks edge sets based on their contribution to the reconstruction of the original graph. Here we use two examples to demonstrate how this process retains important local topological structure during each reconstruction stage. By applying GTI to the BA network, the method learns that there are six stages for reconstruction of the original topology (with stages 1,3, and 5 shown in Figure \ref{evolve-synthetic}). In our second example, we demonstrate GTI reconstruction for the real-world RoadNet dataset. Similarly, Figure \ref{evolve-real} demonstrates the learned reconstruction stages 4,8, and 11.

\textbf{BA Network Stage Analysis:} 
We demonstrate the reconstruction process of a BA network in Figure \ref{evolve-synthetic}, with the top row demonstrating the entire reconstruction process of the full network. We clearly observe that each reconstructed network becomes denser and denser as additional stages are added. The bottom row of Figure \ref{evolve-synthetic} shows the subgraphs corresponding to nodes 0 to 19 at each reconstruction stage. We observe that these subgraphs retain the most important feature of the original subgraph (the star structures at node 0), even during the first reconstruction stage. In addition, we observe that the final stage exactly corresponds to the original stage (again noting as in Figure \ref{fig-compare} that this is extremely difficult when training a single GAN directly on the original graph).


\textbf{Road Network Stages Analysis:} \label{road}
We observe in Table \ref{tab:Table-datasets} that the retained edge percentages of the RoadNet reconstruction decrease more consistently with each stage than in the BA network. This is reasonable, because geographical distance constraints naturally result in fewer super hubs, with each node having less variability in its degree. In Figure \ref{evolve-real}, we observe the reconstruction of the full network, and the node 0 to node 19 subgraph of RoadNet. We observe in the bottom row of Figure \ref{evolve-real} that the dominant cycle structure of the original node 0-19 subgraph clearly emerges.

We also observe an interesting property of the stages of the original graph in the top row of Figure \ref{evolve-real}. As SNAP does not provide the latitude and longitude of nodes, we cannot use physical location. We instead calculate the modularity of each stage, where modularity represents how tight the community is \cite{newman2006modularity} the tighter the community, the larger the modularity). We found that the modularity deceases from 0.98 to 0.92 approximately linearly. This suggests that the GTI stages first prioritizes the clustering of nodes (through edge connections) over connections between clusters. This indicates that GTI views the dense connections between local neighborhoods as a particularly representative topological property of road networks.

\subsection{Global Topological Features}
In the previous section, we demonstrated GTI's ability to the preserve local topological features. Here we demonstrate the ability of GTI reconstruction stages to preserve global topological features, focusing on degree distribution and the distribution of cluster coefficients.

\begin{figure*}[h]
\centering
\scalebox{0.9}{
  	\includegraphics[width=\linewidth]{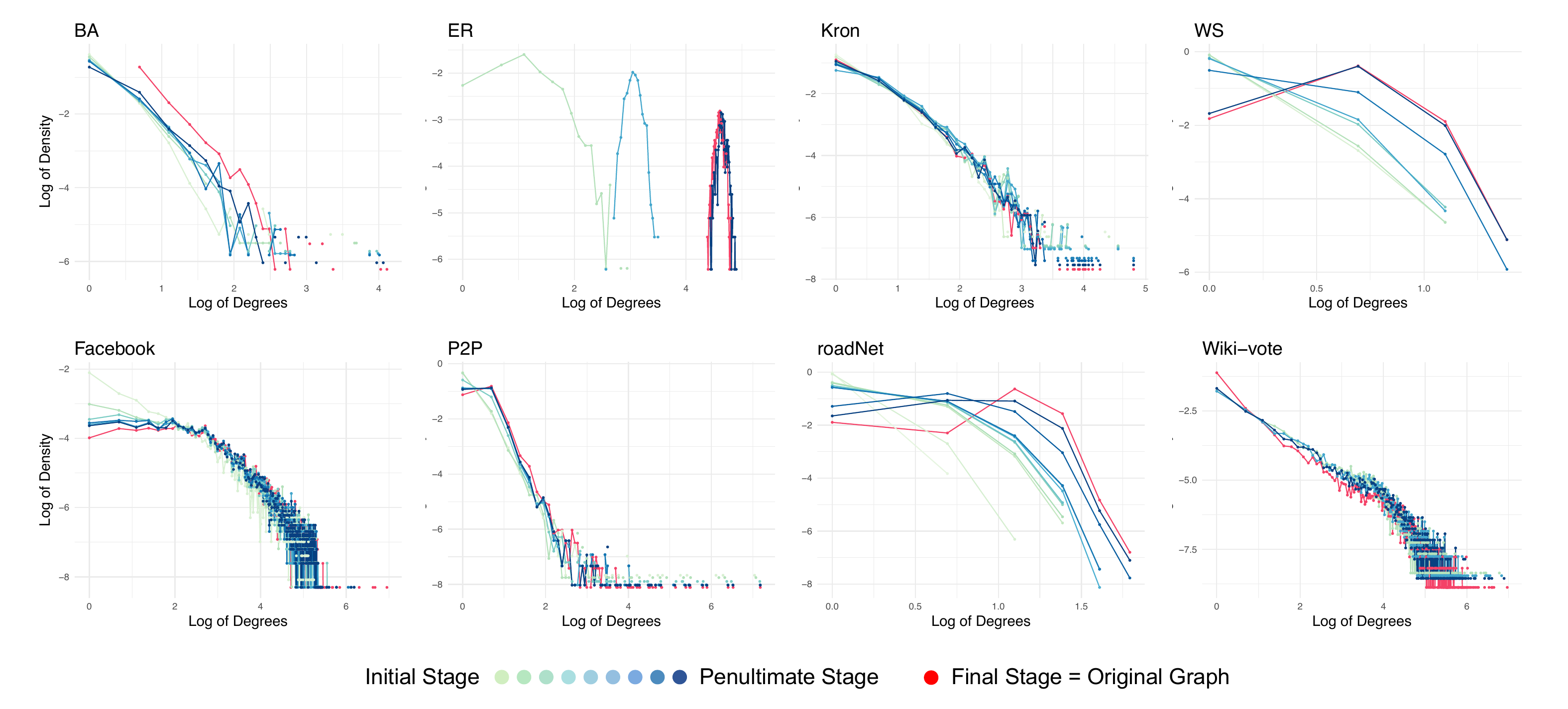}
}
  \caption{Degree distributions for 8 datasets.}
  \label{degree-distribution}
\end{figure*}

\begin{figure*}[h]
\centering
\scalebox{0.9}{
  	\includegraphics[width=\linewidth]{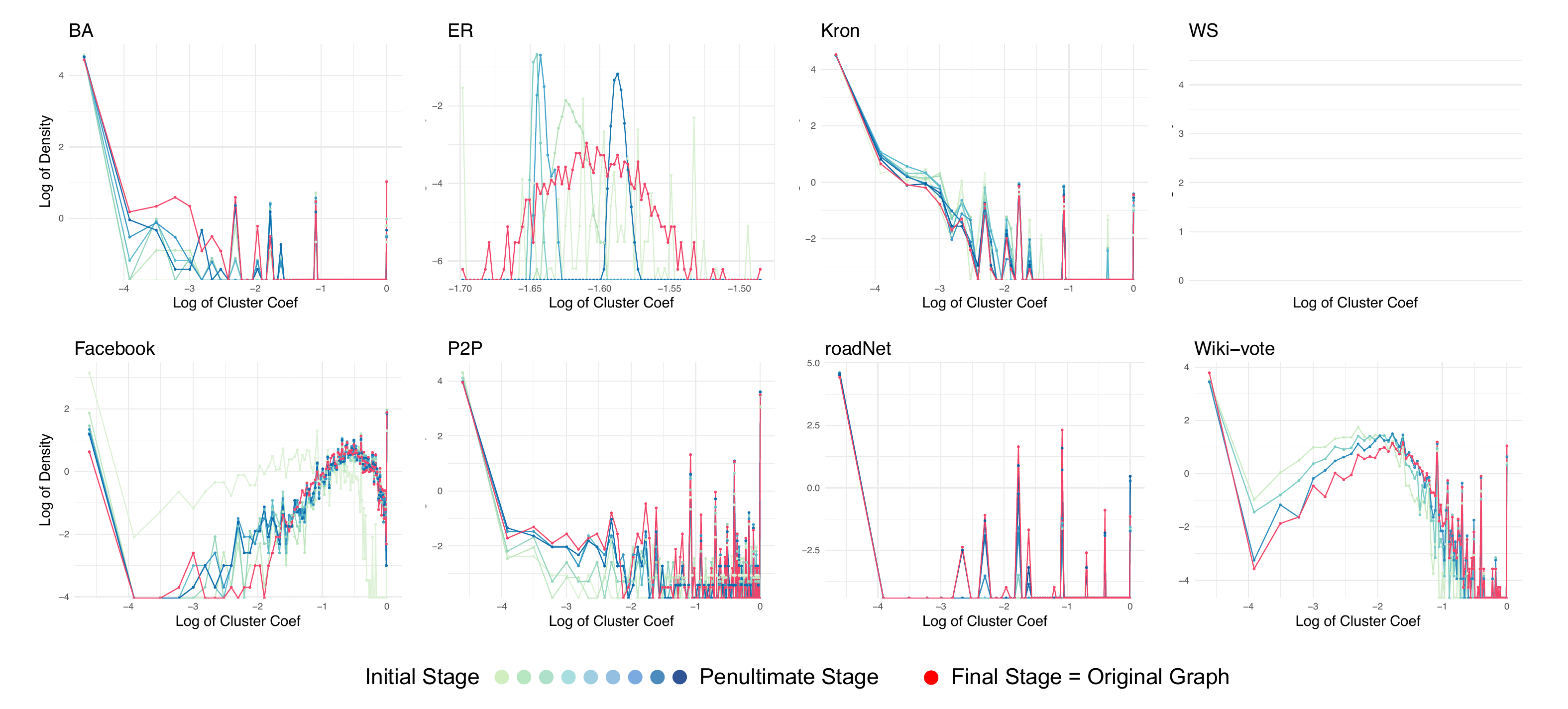}
}
  \caption{Cluster coefficient distributions for 8 datasets.}
  \label{cc-distribution}
\end{figure*}

Figure \ref{degree-distribution} and Figure \ref{cc-distribution} respectively show the log-log degree distributions and log-log cluster coefficient distributions for each of the datasets given in Table \ref{tab:Table-datasets}. 
In Figure \ref{degree-distribution} (\ref{cc-distribution}), the horizontal axis in each degree distribution represents the ordered degrees (ordered cluster coefficient), with the vertical axis representing the density of each degree (the density of each cluster coefficient). The red line is used to demonstrate the degree distribution (cluster coefficient distribution) of the original graph, with the distributions of the ordered stages represented by a color gradient from green to blue. 

We observe that with the exception of the ER network, the degree distributions and the cluster coefficient distributions of early stages are similar to the original graphs, and only become more accurate as we progress through the reconstruction stages. 
Although the degree distributions and cluster coefficient distributions for the early stages of the ER network reconstruction are shifted, we observe that GTI quickly learns the Poisson like shape in degree distribution, and also learns the ``peak-like'' shape in the cluster coefficient distribution. This is particularly noteworthy given that the ER model has no true underlying structure (as graphs are chosen uniformly at random).
Finally, we note that the cluster coefficient of the WS network is zero. For each stage of the WS network, the cluster coefficients are likewise zero. This means GTI learns this feature very quickly, even in the first few stages. We cannot take the log of zero, but include the WS plot in Figure \ref{cc-distribution} for the sake of completeness.

In addition, we also use Frobenius norm \cite{meyer2000matrix} (see Equation \ref{eq-F-norm}) and average node-node similarity \cite{blondel2004measure} (see Equation \ref{eq-nn-sim}) to evaluate the similarity between generated stages and the original graph. Let $A=G-G'_{L-1}$, where $G'_{L-1}$ is the adjacency matrix of the penultimate stage. The notation $A^T$ denotes the matrix transpose of $A$, and $sim\left ( n_{i}^{G'_{L-1}}, n_{j}^{G} \right )$ represents the node-node similarity with mismatch penalty \cite{heymans2003deriving} between node $i \in G'_{L-1}$ and node $j \in G$. $\left | N^G \right |$ is the total number of nodes in original graph. Here, F-norm calculates the distance between two adjacency matrices, and average node-node similarity indicates how $G'_{L-1}$ resembles $G$.

\begin{equation}
F\_norm = \sqrt{\sum_{i=1}^N\sum_{j=1}^Na_{ij}} =\sqrt{trace(A^T A)}
\label{eq-F-norm}
\end{equation}
\begin{equation}
sim(G'_{L-1}, G) = \frac{\sum_{i \in N}\sum_{j \in N} sim\left ( n_{i}^{G'_{L-1}}, n_{j}^{G} \right )}{\left | N^G \right |}
\label{eq-nn-sim}
\end{equation}

For comparison, we use the same model parameters to generate 100 ensembles of BA, ER, Kronecker and WS networks. These newly generated networks serve as base models to help us evaluate the similarity between the penultimate stage and the original graph.
Table \ref{tab:Fdistance} and \ref{tab:avgNode} respectively display the F-norm and the average node-node similarity between the penultimate stage and original graph. Here, bold letters imply best values between the penultimate stage and the mean value of corresponding base models for each dataset. 

\begin{itemize}
    \item Table \ref{tab:Fdistance} shows that 3 of 4 penultimate stages (i.e. BA, Kronecker, WS) have smaller F-norm distance, which means GTI successfully maintains the topological information of the original graph.
    \item Table \ref{tab:avgNode} shows that penultimate stage of BA has larger average node-node similarity with the original graph, and penultimate stages of Kronecker and WS and the corresponding base model have identical similarity to the original graph. These results give a solid evidence that GTI also has the ability to attain good node-level similarity to the original graph.
    \item The ER network is a totally random graph model and hence GTI performs slightly worse than base models (i.e., lack of important topological structure for GTI to learn).
\end{itemize}

\begin{table*}[!t]
  \caption{F-norm distance numerical evaluation on penultimate stage and original graph}
  \label{tab:Fdistance}
  \centering
  \scalebox{1}{
  \begin{tabular}{llllll}
\toprule
Graph     & Penultimate Stage & BaseModel       & BaseModel\_min  & BaseModel\_mean & BaseModel\_max  \\ \midrule
BA        & \textbf{53.0283}           & 62.9031$\pm$0.0859  & 62.8172  & 62.9031  & 62.9762  \\
ER        & 285.8566                   & 282.9252$\pm$0.5637 & 282.3615 & \textbf{282.9252} & 283.4431 \\
kronecker & \textbf{124.9320}           & 125.323$\pm$0.2757  & 125.1987  & 125.1987  & 125.5987  \\
WS        & \textbf{44.4522}            & 44.6094$\pm$0.0448  & 44.5646  & 44.6094  & 44.6542  \\ \bottomrule
  \end{tabular}}
\end{table*}

\begin{table*}[!t]
  \caption{Average node-node similarity numerical evaluation on penultimate stage and original graph}
  \label{tab:avgNode}
  \centering
  \scalebox{1}{
  \begin{tabular}{llllll}
\toprule
Graph     & Penultimate Stage & BaseModel       & BaseModel\_min  & BaseModel\_mean & BaseModel\_max  \\ \midrule
BA        & \textbf{99.9707\%} & 99.9640$\pm$0.0068\% & 99.9573\% & 99.9640\%          & 99.9681\% \\
ER        & 99.6372\%          & 99.6590$\pm$0.0277\% & 99.6338\% & \textbf{99.6590\%} & 99.6867\% \\
kronecker & \textbf{99.2240\%} & 99.2241$\pm$0.0650\% & 99.1606\% & \textbf{99.2240\%} & 99.2889\% \\
WS        & \textbf{99.9994\%} & 99.9994$\pm$0.0001\% & 99.9993\% & \textbf{99.9994\%} & 99.9995\% \\ \bottomrule
  \end{tabular}}
\end{table*}

\subsection{Comparison with Graph Sampling}
The graphs generated by GTI can be considered as samples of the original graph in the sense that they are representative subgraphs of a large input graph. We compare the performance of GTI with that of other widely used graph sampling algorithms (Random Walk, Forest Fire and Random Jump) with respect to the ability to retain topological structures \cite{leskovec2006sampling}. We achieve this by demonstrating subgraph structure on the BA and Facebook datasets, comparing stage 1 of GTI against the graph sampling algorithms (designed to terminate with the same number of nodes as the GTI stage).

\begin{figure}[h]
  	\includegraphics[width=\linewidth]{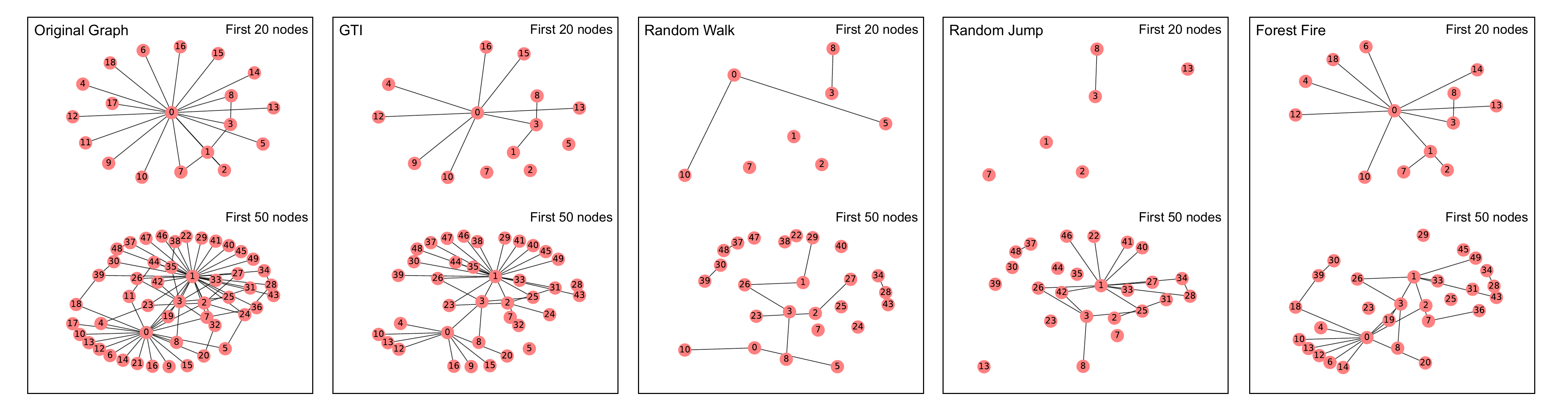}
  \caption{Comparison with graph sampling methods on the BA subgraphs.}
  \label{fig-sampling1}
\end{figure}

\begin{figure}[h]
  	\includegraphics[width=\linewidth]{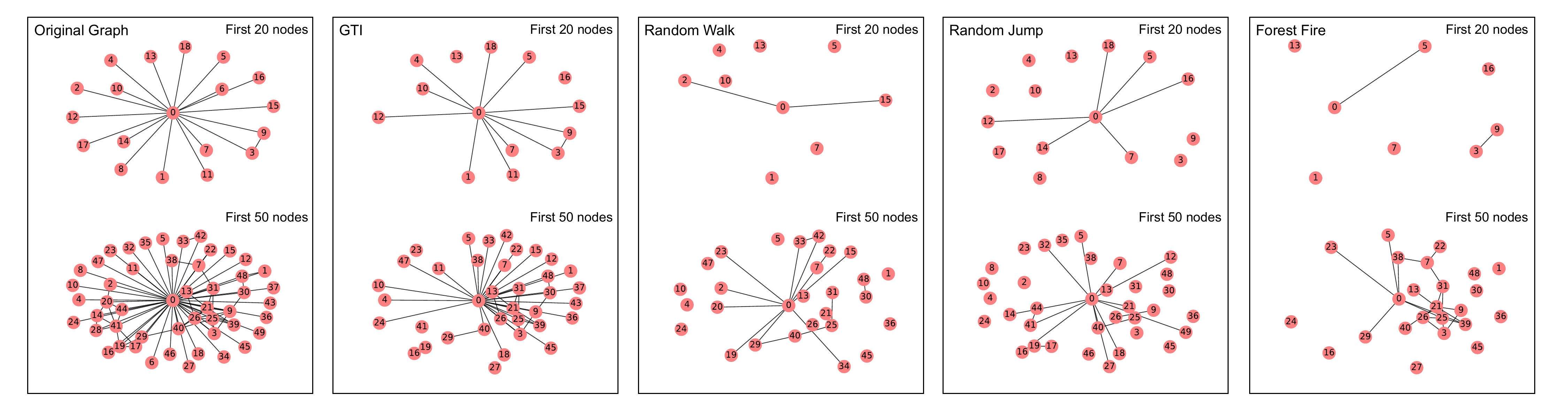}
  \caption{Comparison with graph sampling methods on the Facebook subgraphs.}
  \label{fig-sampling2}
\end{figure}

To avoid messy graph plots, we take out each of subgraphs from BA and Facebook network (nodes 0-19 and nodes 0-49) to visually compare the ability of the first stage of GTI to retain topological features in comparison to the three graph sampling methods.
In Figure \ref{fig-sampling1}, we observe that stage 1 of GTI has retained a similar amount of structure in the 20 node BA subgraph as Forest Fire~ \cite{leskovec2005graphs}, while demonstrating considerably better retention than either Random Walk or Random Jump. However, for 50 node BA subgraph, only GTI has the ability to retain the two super hubs present in the original graph.
In Figure \ref{fig-sampling2}, we observe that GTI demonstrates vastly superior performance to the other methods when run on the Facebook dataset, which has a number of highly dense clusters with very sparse inter-cluster connections.


\section{Discussion and Future Work}

This paper aims to leverage the success of GANs in (unsupervised) image generation to tackle a fundamental challenge in graph topology analysis: a model-agnostic approach for learning graph topological features. By using a GAN for each hierarchical layer of the graph, our method allows us to reconstruct input graph very well, preserving both local and global topological features. 
In addition, our method is able to automatically learn the number of stages required to reconstruct stages and the graph itself non-parametrically. This is potentially advantageous in terms of understanding the distinct stages of graph reconstruction.

Our experimental results show promising results on the capability of GTI for learning distinct topological features from different graphs. To the best of our knowledge, there is not a single graph model that can capture these distinct topological features. A clear direction of future research is in extending the approach to allow the input graph to be directed and weighted, or with node attributions.

\bibliography{paper}
\bibliographystyle{aaai}

\end{document}